\documentclass[letterpaper,11pt]{article}

\usepackage{amsmath,amssymb,textcomp,graphicx,array,natbib} 
\usepackage{fullpage,rotate}
\usepackage[small,bf,up]{caption}

\usepackage{sistyle}
\SIthousandsep{,}

\usepackage{xcolor}

\def\half{\mbox{{\small$\frac{1}{2}$}}}
\def\shalf{\mbox{{\tiny$\frac{1}{2}$}}}

\def\one{\mbox{{\bf 1}}}

\begin{document}
\noindent
{\LARGE \sf Non-Stationary Spatial Modeling}\\[0.5cm]
 Dave Higdon, Jenise Swall, John Kern\\[.6cm]

\noindent
{\small
Standard geostatistical models assume stationarity and rely on a
variogram model to account for the spatial dependence in the observed
data.  In some instances, this assumption that the spatial dependence
structure is constant throughout the sampling region is clearly
violated.  We present a spatial model which allows the spatial
dependence structure to vary as a function of location.  Unlike
previous formulations which do not account for uncertainty in the
specification of this non-stationarity (eg. Sampson and Guttorp
(1992)), we develop a hierarchical model which can incorporate this
uncertainty in the resulting inference.  The non-stationary spatial
dependence is explained through a constructive "process-convolution"
approach, which ensures that the resulting covariance structure is
valid.  We apply this method to an example in toxic waste remediation. \\[.2cm]
}


\section{INTRODUCTION}

Modeling spatial data with Gaussian processes is the common
thread of all geostatistical analyses.
Some notable references in this area include
Matheron (1963),
Journel and Huijbregts (1978),
Ripley (1981),
Isaaks and Srivastava (1990),
and Cressie (1991).
A common approach
is to model spatial dependence through the {\em covariogram} function,
so that covariance between any two points depends only on
the distance between them.  Distance is typically Euclidean, though
other metrics are sometimes used.

Such a specification yields a {\em stationary} random field, which
has a distribution that is invariant under spatial shifts.
If the field is also invariant under rotations, it is called
{\em isotropic}.  Assuming a field to be stationary and
isotropic is natural in some settings, especially when the
spatial region of interest is relatively small.  Though assuming
a stationary model gives sensible results in many applications, 
cases arise when researchers have found it necessary to develop
models that allow for heterogeneous spatial covariance structure
(Haas, 1990; Sampson and Guttorp, 1992; 
Loader and Switzer, 1992; Le and Zidek 1992, 1997).
Of these, only Hass (1990) gives a method for fitting
a non-stationary spatial model using a single realization from
a spatial process, although statements regarding uncertainty
are difficult to obtain under this approach.  
The remaining references above all take advantage of 
repeated measurements over the spatial locations. 

In this paper, we propose an alternative model for accounting
for heterogeneity in the spatial covariance function, which is
based on a moving average specification of a Gaussian process.
Inference is carried out using a hierarchical modeling structure
so that statements regarding uncertainty may be obtained.
A large class of stationary spatial processes may be represented
as a moving average of a Gaussian white noise process -- a white
noise process convolved with some point spread function, or kernel
(see Thi\'ebaux and Pedder, 1987, ch.~5, for example). 
We extend this characterization so that this kernel evolves over
spatial location.  The following section details this new formulation
while Section 3 gives the details of using this specification to 
model spatial data and applying this approach to data obtained from
the Piazza Road Superfund Site.  We finish the paper with a 
brief discussion.

\section{SPECIFYING SPATIAL COVARIANCE}

We now give details on this process convolution approach
for constructing non-stationary
Gaussian processes in two dimensions.  We first consider
stationary formulations obtained through a moving average
specification, and then extend this to allow for processes
with covariance structure that slowly changes over space.
We consider a couple of approaches for specifying a spatially
evolving family of smoothing kernels which, in turn, defines
a spatial covariance function for the process.

\subsection{MOVING AVERAGE PROCESSES}

Any stationary Gaussian process $z(s)$ which has a correlogram
 $\rho(d)$ given by
\[
 \rho(d) = \int_{R^2} k(s)k(s-d) ds
\]
can be expressed as the convolution of a Gaussian white noise 
process $x(s)$ with convolution kernel $k(s)$ 
\[
  z(s) = \int_{R^2} k(s-u) x(u) du.
\]
By a white noise process $x(s)$ we mean a process
for which 
\[
 z_A = \int_A x(u) du \sim N(0,\sigma^2 \mbox{area}(A)) 
       \mbox{ \hspace{.5cm} and \hspace{.5cm}} 
       \mbox{cor}(z_A,z_B) = \mbox{area}(A \cap B),
\]
for any subregions $A$ and $B$ contained in the plane $R^2$.

Primarily for analytical tractability and
computational convenience, we focus on the
two-dimensional standard normal kernel 
\[
  k(s) = \frac{1}{2\pi} \exp\{-\half s^Ts \}
\]
which gives the so-called Gaussian correlation function
in two dimensions
\[
  \rho(d) = \exp\{- d^T d \}.
\]
This isotropic specification can be extended to allow for
rotating as well as shrinking and stretching of the
coordinate axes.  In this case we can represent $k_(s)$ 
as a bivariate normal kernel with covariance matrix $\Sigma$ 
\[
  k(s;\Sigma) = \frac{1}{2\pi} |\Sigma|^{-\shalf} 
                \exp\{-\half s^T \Sigma s \} 
\]
which is isotropic if one uses the transformed 
distance $s' = \Sigma^{-\shalf}s$.  Hence, the three
parameters that make up $\Sigma$ determine the 
anisotropy in the spatial correlation. 

\subsection{SPATIALLY EVOLVING KERNELS}

To account for non-stationarity, we now allow the smoothing
kernel to depend on spatial location $s$.  We
use $k_s(\cdot)$ to denote a kernel which is centered at
the point $s$ and whose shape is a function of location $s$.
Once $k_s(\cdot)$ is specified for all $s \in R^2$, 
the correlation between two points $s$ and $s'$ is then
\begin{equation}
\label{eq:corf}
\rho(s,s') = \int_{R^2} k_s(u) k_{s'}(u) du.
\end{equation}
Note that the spatial correlation function is not 
a correlogram since it is no longer a function
of distance alone. 

Because of the constructive formulation under the moving 
average specification, the resulting correlation function 
$\rho(s,s')$ is certain to be positive definite, and the
resulting non-stationary process
\[
  z(s) = \int_{R^2} k_s(u) x(u) du
\]
is certain to be valid -- no matter how the
family $\{k_s(\cdot)\}_{s\in R^2}$ is chosen, 
provided $\sup \int_{R^2} k_s(u)^2 du < \infty$.  
We favor working with the kernels $k_s(\cdot)$ rather
than working
directly with the correlation function $\rho(s,s')$ 
since this makes it difficult to ensure symmetry and positive definiteness
for all $s$ and $s'$. 
 
Of course, we require that some regularity be injected
into the formulation by forcing the kernels $\{k_s(\cdot)\}$ 
to evolve smoothly over space.  But first, we focus on
the form of the kernels.  For the application presented
in this paper we define each $k_s(\cdot)$ to be a normal
kernel centered at $s$ with spatially varying
covariance matrix $\Sigma(s)$.  In this case if we
parameterize $\Sigma(s)$ and $\Sigma(s')$ by
\[
 \Sigma(s) = 
      \left(
        \begin{array}{rr} a^2 &  \rho ab \\
                          \rho ab &  b^2 \end{array}
      \right)
 \mbox{\hspace{.5cm} and \hspace{.5cm}} 
 \Sigma(s') = 
      \left(
        \begin{array}{rr} {a'}^2 &  \rho' {a'}{b'} \\
                          \rho' {a'}{b'} &  {b'}^2 \end{array}
      \right)
\]
the correlation function is given by the complicated, but easy to
compute formula
\begin{equation}
\label{eq:corf2}
\rho(s,s') = 
  \frac{1}{q_1}\exp\left\{-\frac{1}{q_2} (s-s')^T W (s - s') \right\}
\end{equation}
where 
\begin{eqnarray*}
  W & = & 
      \left(
        \begin{array}{cc} b^2+{b'}^2 & -(\rho ab+  \rho' {a'}{b'}) \\
                 -(\rho ab+  \rho' {a'}{b'})    &  a^2+{a'}^2 \end{array}
      \right) \\
  q_1 & = & 
           2\pi a a' b b'
           \sqrt{(1-\rho^{2})(1-{\rho'}^{2})}\: 
           \sqrt{- \frac{(\rho^{2}-1)b^{2} + 
           ({\rho'}^{2}-1){b'}^{2}}{(\rho^{2}-1)
           ({\rho'}^{2}-1)b^{2}{b'}^{2}}}\: \times \\
      & &
           \sqrt{ \frac{2 \rho {\rho'} a a' b b' +
           a^{2}((\rho^{2}-1)b^{2}-{b'}^{2}) + a'^{2}
           (({\rho'}^{2}-1){b'}^{2})-b^2}{ a^{2}a'^{2}
           ((\rho^{2}-1)b^{2} + ({\rho'}^{2}-1){b'}^{2})}} \\
  q_2 & = &
     {2(2  \rho  {\rho'}  a  {a'}  b  {b'} + 
      a^{2}((\rho^{2}-1)b^{2}-{b'}^{2})-
      {a'}^{2}(({\rho'}^{2}-1){b'}^{2}-b^2))}  
\end{eqnarray*}

In order to assure that the kernels $\{k_s(\cdot)\}$ vary
smoothly over space, we parameterize $\Sigma$, and then
allow the parameters to evolve with location.  We are
currently experimenting with a number of parameterizations,
but for this paper we'll focus on a geometrically based
specification which readily extends beyond the use
of the Gaussian kernels considered here.

There is a one to one mapping from a bivariate normal distribution
to its one standard deviation ellipse, so we define a spatially varying
family of ellipses which, in turn, defines the spatial 
distribution for $\Sigma(s)$.  
Let the two focus points in $R^2$ $(\psi_x,\psi_y)$ and 
$(-\psi_x,-\psi_y)$ define an ellipse centered at the
origin with fixed area $A$.  This then corresponds to
the Gaussian kernel with covariance matrix $\Sigma$ defined by
\[
 \Sigma^{\shalf} = 
      \tau_z
      \left(
        \begin{array}{cc} 
            \left[\frac{\sqrt{4A^2 + ||\psi||^4 \pi^2}}{2\pi} + 
            \frac{||\psi||^2}{2} \right]^{\shalf} & 0 \\
            0 & 
            \left[\frac{\sqrt{4A^2 + ||\psi||^4 \pi^2}}{2\pi} - 
            \frac{||\psi||^2}{2} \right]^{\shalf} \end{array}
      \right) 
      \left(
        \begin{array}{rr} 
          \cos \: \alpha & \sin\: \alpha \\
          -\sin \: \alpha & \cos \: \alpha \end{array}
      \right) 
\]
where $||\psi||^2 = \psi_x^2 + \psi_y^2$,  
$\alpha = \tan^{-1} \psi_y/\psi_x$, and the parameter $\tau_z$ 
serves to scale the kernel. 
This corresponds to a local scaling and rotation at a
given site $s$.
Now we can define a spatial distribution of kernels by
specifying that $\psi_x(s)$ and $\psi_y(s)$ are distributed
independently, according to independent Gaussian fields, each with 
covariance function $\exp\{-(d/\tau_\psi)^2\}$.  The resulting
kernel is then taken to be centered at location $s$. This means
that marginally,  $(\psi_x(s),\psi_y(s))$ has a bivariate 
standard normal distribution for any $s$.  Though any
covariance function could be used here, our preference is for
something simple, and something that ensures very smooth
spatial variation for the ellipses.  The choice of $\tau_\psi$ 
will certainly depend on the application, however the choice
of ellipse area $A$ is something we prefer to fix at a value
that gives sensible ellipses when $(\psi_x(s),\psi_y(s))$ 
are drawn from their Gaussian field distributions.  After a
fair bit of simulation and experimentation, we have settled
on $A = 3.5$.  Note that the scale parameter $\tau_z$ controls
the size of the kernels $\{k_s(\cdot)\}$, and hence serves
to control the range of spatial dependence of the resulting
process $z(s)$.
\begin{figure}[hb]
  \centerline{
   \includegraphics[totalheight=6.0in,angle=-90] {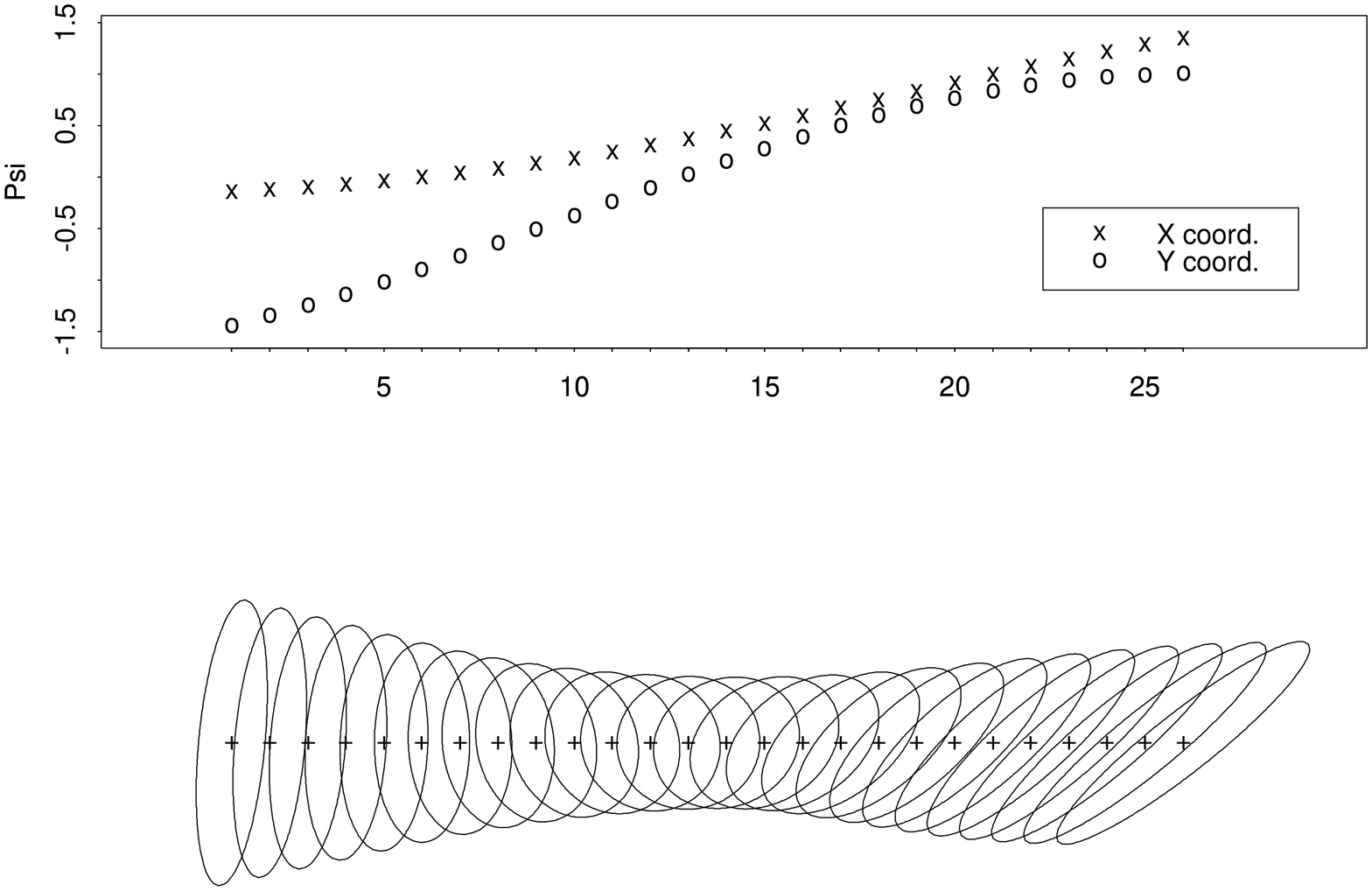}}
   \caption{Ellipses derived 
   from a sequence of foci points $(\psi_x,\psi_y)$. 
   Both the $\psi_x$s and the $\psi_y$s are drawn from separate,
   independent Gaussian processes with covariogram $\exp\{-d/10\}$,
   where $d$ denotes distance.  For each focus $(\psi_x,\psi_y)$,
   the resulting ellipse is plotted directly below it.
 \label{fig:coil}}
\end{figure}

%
Figure 1 shows a spatially evolving sequence of focus 
points $(\psi_x(s),\psi_y(s))$ in the upper frame, and shows
the resulting ellipses in the lower frame.  This demonstrates how the
spatially distributed pairs $(\psi_x(s),\psi_y(s))$ give
rise to a spatially distributed covariance matrix $\Sigma(s)$.
Finally, Figure 2 shows a realization from this non-stationary
random field that has just been specified.  The first frame
shows the ellipses corresponding to the 
realization $\{(\psi_x(s),\psi_y(s))\}$ drawn from their Gaussian
process distributions.  Conditional on these kernels, the 
covariance matrix for a $21 \times 21$ lattice of sites is
constructed and a multivariate normal draw is then made.  The
resulting surface is shown in the second and third frames.
Note that the ellipses shown in Figure 2 are shrunken down
by a factor of 10 so that they can be seen on the plot.
\begin{figure}
  \centerline{
   \includegraphics[totalheight=6.3in,angle=-90] {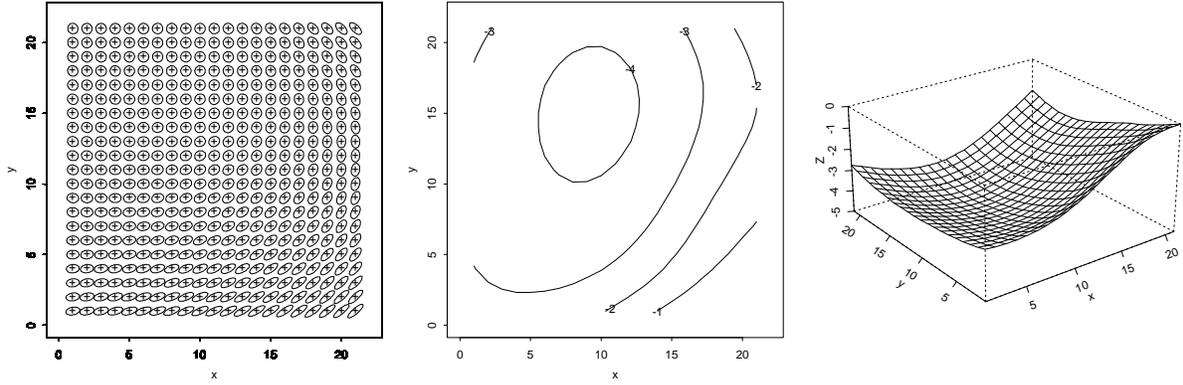}}
   \caption{A  single realization from
   a process convolution process with spatially
   evolving kernels.  The ellipses depict one standard deviation
   ellipses (shrunken by a factor of 10) for the 
   kernel $k_s(\cdot)$ associated with that location.
   Conditional on the kernels, a surface is generated; 
   2b and 2c show the contour plot and perspective plot
   of this realization.
 \label{fig:triple}}
\end{figure}


\section{APPLICATION IN ENVIRONMENTAL MONITORING}

This non-stationary specification is now used as a spatial
prior distribution in
estimating the spatial distribution of
dioxin concentrations on the Piazza Road pilot study
area, which is part of an EPA Superfund site in Missouri.  Here dioxin
originally applied to the road was transported through a small stream
channel at the site.  Hence, the concentrations are much higher
at locations near the stream channel.  Also, it is expected that the
spatial dependence should be stronger along the direction of
the channel, which does not follow a straight path.  Hence this
is a case where allowing the spatial dependence to vary with
location may lead to a more appropriate model.

The actual pilot data contain over 1000 concentration measurements,
but we use only a subset of $n=60$ data points, restricted over a subset of
the pilot area as shown in Figure 3.  
We use $s_1,\ldots,s_n$ to denote the
locations and $y=(y_1,\ldots,y_n)^T$ to denote the log
of the concentration measurements 
corresponding to the $n$ locations.
We assume the log of the dioxin concentration
readings $y$ can be decomposed into an overall mean $\mu$, 
spatial trend $z(s)$, and an i.i.d.~error component $\epsilon$ 
so that
\[
  y = \mu\one + z + \epsilon.
\]
We assume the errors have an i.i.d.~Gaussian distribution with
precision $\lambda_y$.  The formulation is completed by the prior
specifications for $z$ and $\lambda_y$.  For $\lambda_y$ we use
a gamma prior; for $z$ we use the non-stationary Gaussian process
detailed in the previous section.   We give the prior
specification below. 
\begin{eqnarray*}
\lambda_y & \sim & \Gamma(a_y, b_y) \\
\mu & \sim & U(-\infty,\infty) \\
z|\lambda_z,\tau_z,\psi & \sim & 
  N\left(0,\frac{1}{\lambda_z} R_z(\tau_z,\psi) \right) \\ 
\lambda_z & \sim & \Gamma(a_z, b_z) \\
\tau_z & \sim & U(\ell_z, u_z) \\
\psi_x|\tau_\psi &\sim& N(0,R_\psi(\tau_\psi) ) \\
\psi_y|\tau_\psi &\sim& N(0,R_\psi(\tau_\psi) ) \\
\tau_\psi & \sim & U(\ell_\psi, u_\psi) 
\end{eqnarray*}
The hyperparameters $a_y, b_y, a_z$, and $b_z$ are chosen
to give diffuse, but proper Gamma distributions; the correlation
matrix $R_z(\tau_z,\psi)$ is determined by the correlation
function (\ref{eq:corf2}); the correlation matricies $R_\psi(\tau_\psi)$ for 
$\psi_x$ and $\psi_y$ are identical and determined by the 
correlation function $\rho(s,s') = \exp\{-[||s-s'||/\tau_\psi]^2\}$.
Finally, the spatial range parameters $\tau_z$ and $\tau_\psi$ 
are constrained so that the resulting posterior distribution is
proper.  In both cases, we set $\ell=3$ and $u=200$ so that any
plausible value for these parameters is well within this range.

\begin{figure}
  \centerline{
   \includegraphics[totalheight=4.3in,angle=-90] {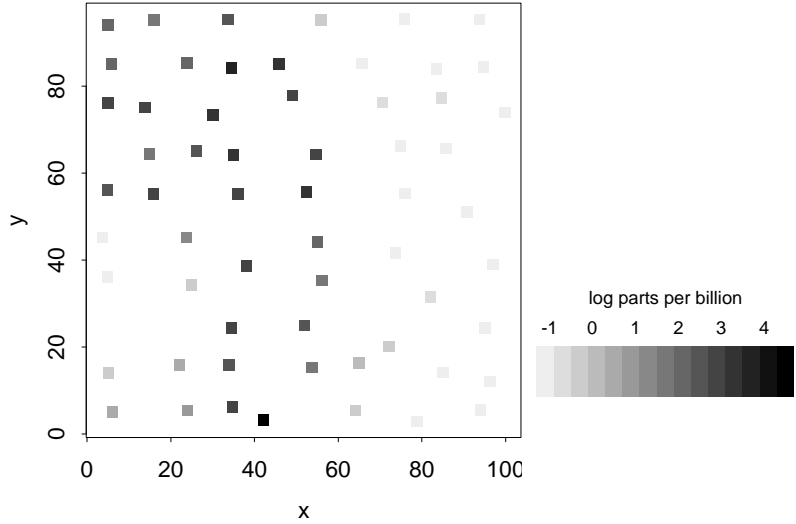}}
   \caption{Data from the
   Piazza Road pilot study.
 \label{fig:piazza}}
\end{figure}


The posterior is explored via MCMC after integrating out
the $z$ term.   For most of the parameters, the full conditionals
are not of a simple form, so stochastic simulation from the
posterior is carried out primarily via Metropolis and Metropolis-Hastings
steps.

\subsection{RESULTS}

Though the MCMC output allows one to explore many facets
of the posterior distribution, we focus here on aspects pertaining
to parameters which control the non-stationarity within 
this application.  The top
row of Figure 4 depicts the spatially distributed
kernel estimates $\{ k_s(\cdot)\}$ at each of the data locations
for $\tau_\psi = 25$ and 50; $\tau_\psi$ is the parameter which controls
the amount of non-stationarity in the spatial correlation for $z(s)$.
Each ellipse corresponds to a one standard deviation ellipse for 
the bivariate kernel $k_s(\cdot)$; each ellipse is scaled down by a factor
of 4 so that it fits into the plotting region.  
These estimates were obtained by maximizing over $\psi_x(s)$ and 
$\psi_y(s)$ while holding the remaining parameters at their posterior
mean estimates.  As expected, when $\tau_\psi = 25$ feet, 
a relatively small value,  a fair bit of non-stationarity is 
present.  As $\tau_\psi$ increases, the foci estimates 
$(\widehat{\psi}_x(s),\widehat{\psi}_y(s))$ become increasingly
uniform across locations.  At $\tau_\psi = 50$, the solution is
effectively the same as in the stationary case.  
The resulting posterior distribution gives a 95\% credible 
interval for $\tau_\psi$ of $(31,36)$.  
Hence the resulting inference for $z(s)$ not only 
allows for non-stationarity, but also accounts for uncertainty
in the degree of the departure from stationarity.
Posterior mean estimates of $\{ k_s(\cdot)\}$ at each of the data locations
are depicted in the bottom rows of Figure 4.  The bottom left
figure shows the ellipses corresponding to the posterior mean of
each pair of foci points.  The bottom right figure shows the 
posterior mean of the actual one sd ellipses themselves where
the ellipses from each MCMC realization are averaged radially.
Hence, the resulting shapes are not ellipses in general.
Note the posterior mean estimates show less regularity than
their modal counterparts in the top row.
\begin{figure}
  \centerline{
   \includegraphics[totalheight=6.3in,angle=-90] {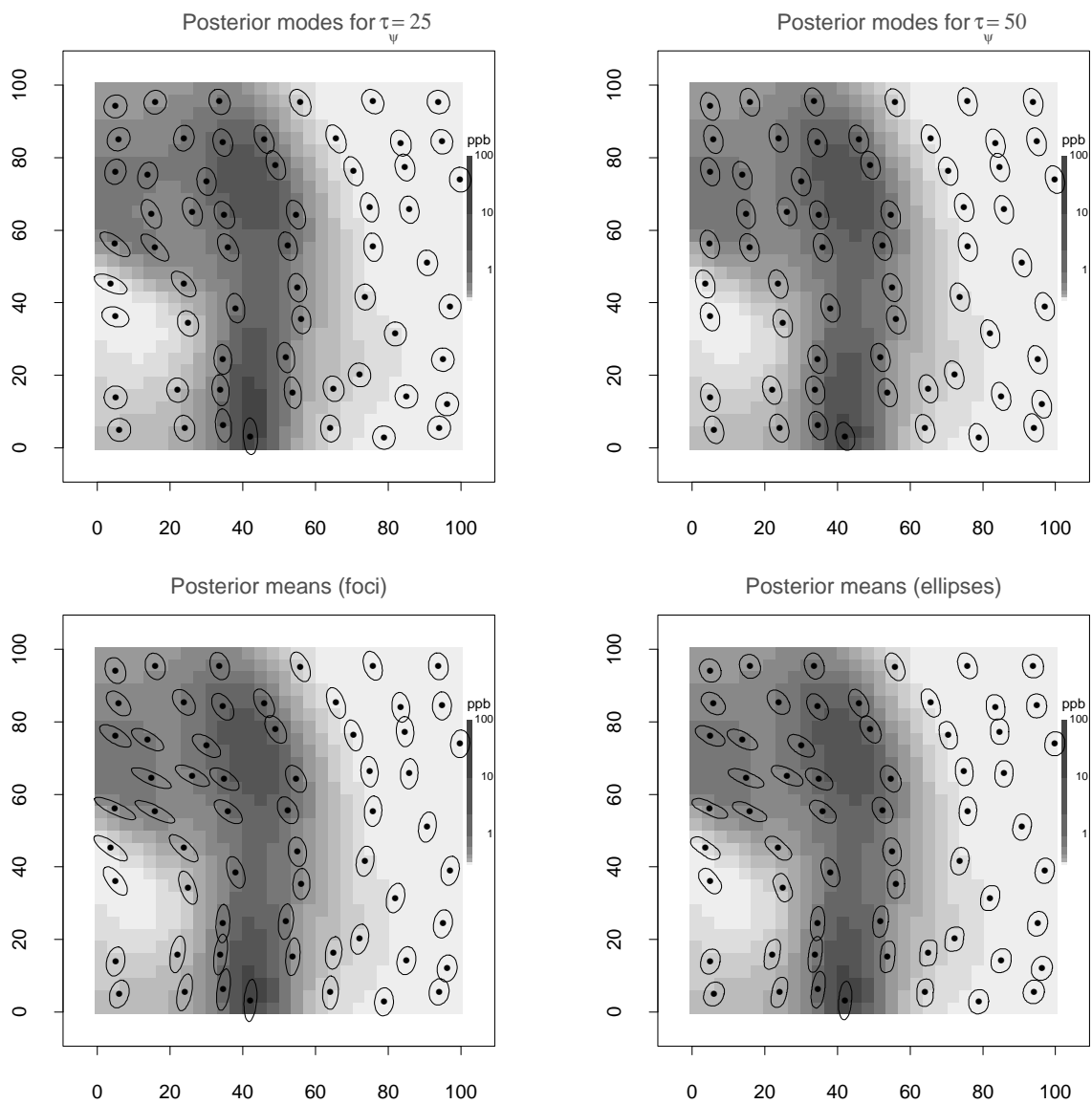}}
   \caption{Top row: 
   Estimated kernels $k_s(\cdot)$ for the Piazza Road data 
   with the range parameter $\tau_\psi$ fixed at 25 and 50.  
   The ellipses are one standard
   deviation ellipses of the bivariate Gaussian kernel at 
   each data point $s$, scaled down by a factor of 4.  
   Bottom left: Ellipses corresponding
   to the posterior means of the foci when 
   $\tau_\psi$ is treated as a parameter.  
   Bottom right: posterior mean of the
   ellipses, obtained by radially averaging the ellipses 
   throughout the simulation. 
   The posterior mean estimate for $\tau_\psi$ is 36.
   The images show the corresponding posterior mean 
   estimates for the dioxin concentrations.
 \label{fig:vfits}}
\end{figure}


\section{DISCUSSION}

We have developed a spatial process model 
which allows for non-stationary spatial dependence and
have also given an application where MCMC is a practical
and effective means of carrying out the estimation.  
The example presented here is unique in that it accounts for 
uncertainty in the extent of the non-stationarity, and
it does not require repeated realizations from the
spatial process.  
Our belief is that incorporating uncertainty due to 
estimating the spatial heterogeneity
is important in determining whether such model extensions
are worthwhile.
Of course, model development is still in
its initial stages and just how effective such a modeling
approach may be in serious applications remains to be tested.
However the initial indications are promising.

Finally, we note that as
with any hierarchical model specification, alternative
prior formulations are possible.  This is particularly true
here since the nature of the models do not give rise to
any inviting conjugate formulations.  The spatial range
parameters $\tau_z$ and $\tau_\psi$ are especially awkward
and our use of uniform prior over a fixed interval is
somewhat clumsy.  However the chosen intervals cover
any remotely plausible range value given the application
considered here.  As for the spatial prior governing the $\psi_x(s)$ and
 $\psi_y(s)$ processes, our preference is for a 
rather simple model.  One plausible alternative would be to
shrink the $\psi$s towards a common $(\psi_x^0,\psi_y^0)$ 
rather than towards the origin.  However, we feel that anything more 
elaborate is not likely to be fruitful since the data 
are unlikely to contain any detailed information about the structure
of the $\psi$s.  Investigations regarding 
prior sensitivity of this approach are currently being conducted. 

\section{ACKNOWLEDGEMENTS}

This research originally began while DH was a research fellow
at the National Institute of Statistical Sciences.  DH would like
to thank Jerry Sacks for his insight and support during the earlier
stages of the research.  Thanks also to Robert Wolpert and
Katja Ickstadt for many helpful discussions.  The research was
supported by National Science Foundation grant DMS-9704425.

\section{}{REFERENCES}

\begin{list}{}{\leftmargin=1em\itemindent=-\leftmargin}

\item
Barry, R.~P. and {Ver Hoef}, J.  (1996).
\newblock Blackbox kriging: spatial prediction without specifying variogram
  models, {\it J. Ag. Bio. and Eco. Statist.}
  {\bf 3}:~1--25.

\item %
Besag, J., Green, P.~J., Higdon, D.~M. and Mengersen, K.  (1995).
\newblock {Bayesian} computation and stochastic systems (with
discussion), {\it Statistical Science} {10}:~3--66. 

\item %
Brown, P.~J., Le, N.~D. and Zidek, J.~V.  (1994).
\newblock Multivariate spatial interpolation and exposure to air pollutants,
{\it Canadian Journal of Statistics} {22}:~489--509.

\item
Cressie, N. A.~C.  (1991).
\newblock {\it Statistics for Spatial Data}, Wiley-Interscience.

\item %
Diggle, P.~J., Moyeed, R.~A. and Tawn, J.~A.  (1998).
\newblock Non-{G}aussian geostatistics, to appear {\it Journal of the Royal Statistical Society (Series B)}.

\item
Guttorp, P. and Sampson P.~D. (1994).
\newblock Methods for estimating heterogeneous spatial covariance
  functions with environmental applications, {\it in} G.~P. Patil and
  C.~R.~Rao (eds), {\it Handbook of Statistics, Vol. 12},
  Elsevier.

\item
Guttorp, P., Meiring, W. and Sampson, P.~D.  (1994).
\newblock A space-time analysis of ground-level ozone data, {\it
  Environmetrics} {\bf 5}:~241--254.

\item %
Haas, T.~C.  (1992).
\newblock Lognormal and moving window methods of estimating acid decomposition,
{\it Journal of the American Statistical Association} {85}:~950--963.

\item
 Higdon, D. (1998). A process-convolution approach to modeling temperatures
         in the North Atlantic Ocean,
         {\it J. Env. and Eco. Statist} {\bf 5}:~173--190.

\item
Isaaks, E.~H. and Srivastava, R.~M.  (1990).
\newblock {\it An Introduction to Applied Geostatistics}, Oxford University
  Press.

\item
Journel, A.~G. and Huijbregts, C.~J.  (1979).
\newblock {\it Mining Geostatistics}, Academic Press.

\item %
Le, N.~D. and Zidek, J.~V.  (1992).
\newblock Interpolation with uncertain spatial covariance,
{\it Journal of Multivariate Analysis} {43}:~351--374.

\item %
Le, N.~D. and Zidek, J.~V.  (1997).
\newblock {B}ayesian multivariate spatial interpolation with missing data by
  design, {\it Journal of the Royal Statistical Society (Series B)} {59}:~501--510.

\item
Loader, C. and Switzer, P.  (1992).
\newblock Spatial covariance estimation for monitoring data, {\it Statist. in
   Env. \& Earth Sci.}, pp.~52-- 70.

\item
Mat\'ern, B.  (1986).
\newblock {\it Spatial Variation (Second Edition)}, Springer-Verlag.

\item
Matheron, G.  (1963).
\newblock Principles of geostatistics, {\it Economic Geology} {\bf
  58}:~1246--1266.

\item
Meiring, W., Monsetiez, P., Sampson, P.~D. and Guttorp, P.  (1996).
\newblock Developments in the modelling of nonstationary spatial covariance
  structure from space-time monitoring data, {\it in} E.~Y. Baafi and
  N.~Schofield (eds), {\it Proc. of the 5th International Geostatist.
  Congress}, Kluwer.

\item
De Olivera, V., Dedem, B. and Short, D.~A. (1997).
\newblock Bayesian prediction of transformed Gaussian random fields, 
  {\it Journal of the American Statistical AssociationJournal of the American Statistical Association} {92}:~1422--1433.

\item
Ripley, B.~D.  (1981).
\newblock {\it Spatial Statistics}, John Wiley \& Sons.

\item
Ryti, R.~T.  (1993).
\newblock Superfund soil cleanup: developing the {P}iazza {R}oad remedial
  design, {\it J. of the Air and Waste Mgmt. Assoc.} {\bf
  24}:~381--391.

\item
Ryti, R.~T., Neptune, B. and Groskinsky, B.  (1992).
\newblock Superfund soil cleanup: applying the {P}iazza {R}oad remediation
  plan, {\it Environmental Testing and Analysis} {\bf 1}:~26--31.

\item %
Sampson, P.~D. and Guttorp, P.  (1992).
\newblock Nonparametric estimation of nonstationary spatial covariance
  structure, {\it Journal of the American Statistical Association}  {87}:~108-119.

\item %
Sampson, P.~D., Guttorp, P. and Meiring, W.  (1994).
\newblock Spatio-temporal analysis of regional ozone data for operational
  evaluation of an air quality model, {\it Proc. of the American
  Statistical Association}.

\item
Thi\'ebaux, H.~J. and Pedder, M.~A.  (1987).
\newblock {\it Spatial Objective Analysis: with Applications in Atmospheric
  Science}, Academic Press, San Diego.

\item %
Tierney, L.  (1994).
\newblock Markov chains for exploring posterior distributions (with
  discussion), {\it Applied Statistics} {21}:~1701--1762.

\item %
{Ver Hoef}, J. and Barry, R.~P.  (1998).
\newblock Constructing and fitting models for cokriging and multivariable
  spatial prediction, to appear {\it Journal of Statistical Planning and Inference}. 

\item %
Wolpert, R.~L. and Ickstadt, K.  (1998).
\newblock Gamma/{P}oisson random field models for spatial statistics, 
  {\it Biometrika} {85}:251-267.

\end{list}

\end{document}